\documentclass[12pt]{article}
\usepackage{amsfonts}

\newcommand{\beq}[1]{\begin{equation}\label{#1}}
\newcommand{\eeq}{\end{equation}}
\newcommand{\bear}[1]{\begin{eqnarray}\label{#1}}
\newcommand{\ear}{\end{eqnarray}}

\renewcommand{\theequation}{\arabic{section}.\arabic{equation}}
\catcode`\@=11 \@addtoreset{equation}{section}\catcode`\@=12

\newcommand{\R}{ {\mathbb R} }

\newcommand{\fnm}{\footnotemark}
\newcommand{\fnt}{\footnotetext}

\begin{document}

 \begin{center}
 \large\bf

 On multitemporal generalization of Newton's gravitational law

 \vspace{15pt}

 \normalsize\bf V.D. Ivashchuk\fnm[1]\fnt[1]{ivashchuk@mail.ru}

 \vspace{5pt}

 \it Center for Gravitation and Fundamental Metrology,
 VNIIMS, 46 Ozyornaya ul., Moscow 119361, Russia  \\

 Institute of Gravitation and Cosmology,
 Peoples' Friendship University of Russia,
 6 Miklukho-Maklaya ul.,  Moscow 117198, Russia \\

 \end{center}

 \vspace{15pt}

 \begin{abstract}

 A $n$-time generalization of Newton's law
 (of universal gravitation) formula
 in $N =n + d + 1$-dimensional  space-time is conjectured.
 This formula  implies a relation for effective $N$-dimensional
 gravitational constant  $G_{eff}  =   G  cos^2 \theta$,
 where $\theta$ is the angle
 between the direction of motion of two particles in
 $n$-dimensional time manifold $\R^n$.

 \end{abstract}

 \section{Introduction}

   In \cite{I-07}  a $n$-time  generalization of the
  Newton's gravitational law formula in the $N= n+1+d$-dimensional space-time
  was suggested.  The generalized Newton's law force reads
   \beq{1.1}
     F =   G   \frac{m_1 m_2}{R^{d}} cos^2 \theta,
   \eeq
  where $G$ is N-dimensional gravitational constant,
  $m_1$ and $m_2$ are masses of interacting point-like particles
  and $R$ is the distance between them (that is supposed
  to be large enough).   Here $\theta$ is the angle
  between the direction of motion of these two particles in
  $n$-dimensional time manifold $\R^n$, or,  more explicitly,
   \beq{1.2}
     \cos \theta = \vec{n}_1 \vec{n}_2,
   \eeq
   where  $n$-dimensional unit vectors $\vec{n}_1$, $\vec{n}_1$
   define the  motion of two particles in the
   time manifold $\R^n$
   \beq{1.3}
     \vec{t}_{1} (\tau) = \vec{n}_1 \tau + \vec{t}_{01},
   \quad   \vec{t}_2 (\tau) = \vec{n}_2 \tau + \vec{t}_{02}.
   \eeq

 \section{The ``derivation''}

  The formula (\ref{1.1})  may be  obtained  using the following
  relation for gravitational potential energy
  \beq{1.4}
  V =   - \bar G \frac{ tr(M_{1} M_{2})}{R^{d-1}},
  \eeq
  where  $\bar G = G/(d -1)$, $M_{1} = (m_1 n^{i}_1 n^{j}_1)$ and
  $M_{2} = (m_2 n^{i}_2 n^{j}_2)$ are (multitemporal)
  mass matrices (inertial or gravitational ones) for the first
  and the second particles, respectively.
  (For $d = 2$ case see \cite{IM-mt2}.)

  Relation  (\ref{1.4}) is
    a special case of multitemporal analogue of
  Newton's law  obtained in  \cite{IM-mt1}
   \beq{1.5}
  V =  - \bar G \frac{ tr(M_{1G} M_{2I})}{R^{d-1}},
   \eeq
  where  $M_{1G}$ is a gravitational mass matrix of the first particle
  and  $M_{2I}$ an inertial mass matrix of the second particle.

  The formula  (\ref{1.5}) was derived in \cite{IM-mt1}
  by considering the motion of the relativistic massive particle
  in a gravitational background described by multitemporal
  generalization  of the Tangherlini solution \cite{T,MP}
  (see \cite{IM-mt1} and Appendix). The derivation of (\ref{1.5}) was
  performed in assumption of a non-relativistic motion
  of the particle at large distances from the
  gravitating center. Recall that the inertial mass
  matrix for the relativistic particle of scalar mass $m$
  is $M = (m n^{i} n^{j})$, where the unit vector $\vec{n} = (n^i)$
  describes the motion in  the time manifold $\R^n$:
  $\vec{t} (\tau) = \vec{n} \tau + \vec{t}_{0}$, where
  $\tau$ is a synchronous time parameter.

  The solution in \cite{IM-mt1} contains both black hole
  configurations and naked singularities as well. The
  relation for  $M_{1G}$ in \cite{IM-mt1} contains
  some extra parameters, but for black hole configurations
  in a space-time with extra time dimensions
  (that are trivial generalization of Tangherlini solution
  \cite{T}) we get that  the gravitational and inertial mass
  matrices are coinciding,  i.e.
   \beq{1.6}
   M_{iG}=  M_{iI},
    \eeq
   $i = 1,2$ (see Appendix).

\section{Discussions}

The relation (\ref{1.1}) implies the following value for the
effective gravitational constant
    \beq{1.7}
     G_{eff}  =   G  cos^2 \theta,
    \eeq
where $\theta = \theta_{i j}$ refers to a pair of two particles
labelled by  indices $i$ and $j$.

It follows from (\ref{1.7}) that $G_{eff} = 0$ for $\theta =
\pi/2$, i.e. the "Newtonian" interaction is absent in this case.
It is important that
   \beq{1.8}
     G_{eff}  \geq 0,
    \eeq
i.e. the repulsion is absent.

Now we put $d = 2$.  Let us consider a flow of particles  moving
in the time manifold $\R^n$ near the $t^1$-axis with small enough
 values of  $\theta_{i j} = \theta_{i j}(t^1)$.
 Then, for proper restrictions on
 $\theta_{i j}$ and  $\dot \theta_{i j}$ imposed, the effective
4-dimensional gravitational constant $G_{eff}$ may have a small
enough variation in $t^1$-time (compatible with the observational
data) and the weak equivalence principle will be not violated.

 \begin{center}
 {\bf Acknowledgments}
 \end{center}

 This work was supported in part by the Russian Foundation for
 Basic Research grants Nr. 09-02-00677-a.

 \renewcommand{\theequation}{\Alph{section}.\arabic{equation}}
 \renewcommand{\thesection}{}
 \setcounter{section}{0}

%%%%%%%%%%%%%%%%%%%%%%%%%%%%%%%%%%%%%%%%%%%%%%%%%%%%%%%%%%%%%%%%
 \section{Appendix}
%%%%%%%%%%%%%%%%%%%%%%%%%%%%%%%%%%%%%%%%%%%%%%%%%%%%%%%%%%%%%%%%

 The solution  in \cite{IM-mt1} reads

 \begin{eqnarray}
 \label{a.1}
  g=&& - \sum_{i=1}^{n} (1 - B R^{1-d})^{a_{i}} dt^{i} \otimes dt^{i} \\
  \nonumber
  && + (1 - B R^{1-d})^{b-1}  dR \otimes dR +
(1 - B R^{1-d})^{b} R^{2} d \Omega^{2}_{d},
 \end{eqnarray}
 where
 \begin{equation}
 \label{a.2}
 b = (1 - \sum_{i=1}^{n} a_{i})/(d-1)
 \end{equation}
 and the parameters $a_{1}, \ldots , a_{n}$ satisfy the relations
 \begin{equation}
 \label{a.3}
 (\sum_{i=1}^{n} a_{i})^{2} + (d-1) \sum_{i=1}^{n} a_{i}^{2} = d.
 \end{equation}
  Here $B$ is constant.
  This solution  is a special case
  of the solution obtained earlier in  \cite{FIM}.

 The black hole configuration takes place when
 the set $(a_{1}, \ldots , a_{n})$ is  coinciding with
 one of the points
 \begin{equation}
 \label{a.4}
 (1,0, \ldots , 0), \ldots ,  (0, \ldots, 0,1).
 \end{equation}

In other cases we get naked singularities \cite{IM-mt1}.

The gravitational mass matrix corresponding to (\ref{a.1})
 was defined in  \cite{IM-mt1} as follows

 \beq{a.5}
 M_{ij} = a_{i} \frac{\delta_{ij} B}{2G}.
 \eeq

 More general form of solution may be obtained by a $O(n)$-rotation
 of time axes $t^1, ..., t^n$. For the set $(1,0, \ldots , 0)$
 we get a ``pure state'' (projector) matrix in (\ref{a.5})
 which may be reduced to the general form $M = (m n^{i} n^{j})$
(where $\vec{n} = (n^i)$ is a vector of unit length)
 by a suitable $O(n)$-rotation.

 \end{document}